\documentclass{aa}  

\usepackage{longtable}
\usepackage{graphicx}
\usepackage{color}

\usepackage{txfonts}
\begin{document} 

   \title{\textbf{
   Transport of the magnetic flux away from a decaying sunspot via convective motions }}
   
    \author{ Chenxi Zheng\inst{1} 
             \and 
             Thierry Roudier \inst{2} 
             \and
             Brigitte Schmieder \inst{3,4} 
             \and  
             Guiping Ruan \inst{1} 
             \and 
             Jean-Marie Malherbe \inst{4}
             \and
             Yang Liu\inst{1}
             \and
            Yao Chen\inst{1}
             \and            
             Wenda Cao \inst{5,6}
             }
\institute{Shandong Provincial Key Laboratory of Optical Astronomy and Solar-Terrestrial Environment, and Institute of Space Sciences, Shandong University, Weihai 264209, China\\
            \email{rgp@sdu.edu.cn}
            \and
             Institut de Recherche en Astrophysique et Plan\'etologie (IRAP), Universit\'e de Toulouse, CNRS, UPS,
      CNES, 14 avenue Edouard Belin, 31400 Toulouse, France
            \and 
             Centre for Mathematical Plasma-Astrophysics, Department of Mathematics, KU Leuven, Celestijnenlaan 200B, 3001 Leuven, Belgium 
             \and 
             LESIA, Observatoire de Paris, Universit\'e PSL, CNRS, Sorbonne Universit\'e, Universit\'e de Paris, 5 place Jules Janssen, 92190 Meudon, France
             \and 
                Center for Solar-Terrestrial Research, New Jersey Institute of Technology, 323 Martin Luther King Blvd., Newark, NJ 07102, USA
            \and    
            Big Bear Solar Observatory, 40386 North Shore Lane, Big Bear City, CA 92316, USA
   }

 
  \abstract
  {The interaction between magnetic fields and convection in sunspots during their decay process remains poorly understood, 
  whereas the formation of sunspots is relatively well studied and fully modeled.
  Works on the velocity scales at the solar surface have pointed to the existence of the family of granules, whose interaction with the magnetic field leads to the formation of supergranules and their networks, which are visible at the solar surface.}  
   {The aim of this paper is to consider relationship between the decay of sunspots and  convection via the motion of the family of granules and   how the diffusion mechanism of  magnetic field  operates in a  decaying sunspot.
}
    {We report the decay of  a sunspot observed by the 1.6m Goode Solar Telescope (GST) with  the TiO Broadband Filter Imager (BFI) and the Near-InfraRed Imaging Spectropolarimeter (NIRIS). The analysis was aided by the Helioseismic and Magnetic Imager (HMI) on board the Solar Dynamic Observatory (SDO). In the first step,  we followed the decay of the sunspot with HMI data  over three days by constructing its evolving area and total magnetic flux. In the second step, the high spatial and temporal resolution of the GST instruments allowed us to analyze the causes of the decay of  the sunspot. 
 Afterward, we followed the emergence of granules in the moat region around  the sunspot over six hours.  The evolution of  the trees of fragmenting granules (TFGs)  was derived based on their relationship with    the horizontal surface flows.}
   {We find that the area and total magnetic flux display an exponential  
   decrease over the course of the sunspot  decay. We identified 22 moving magnetic features (MMFs) in the moats of pores, which is a signature of sunspot decay through diffusion.
   We note that the MMFs were constrained to follow the borders 
   of TFGs  during their journey away from the sunspot. }
  {The TFGs and their development contribute to the diffusion of  the magnetic field outside the sunspot. The conclusion of our analysis shows the important role of the TFGs in sunspot decay. Finally, the  family of granules evacuates the magnetic field.  }


\maketitle

%
\begin{figure*}[!htbp]
\begin{minipage}{\textwidth}
\centering
\includegraphics[width=180mm,angle=0,clip]{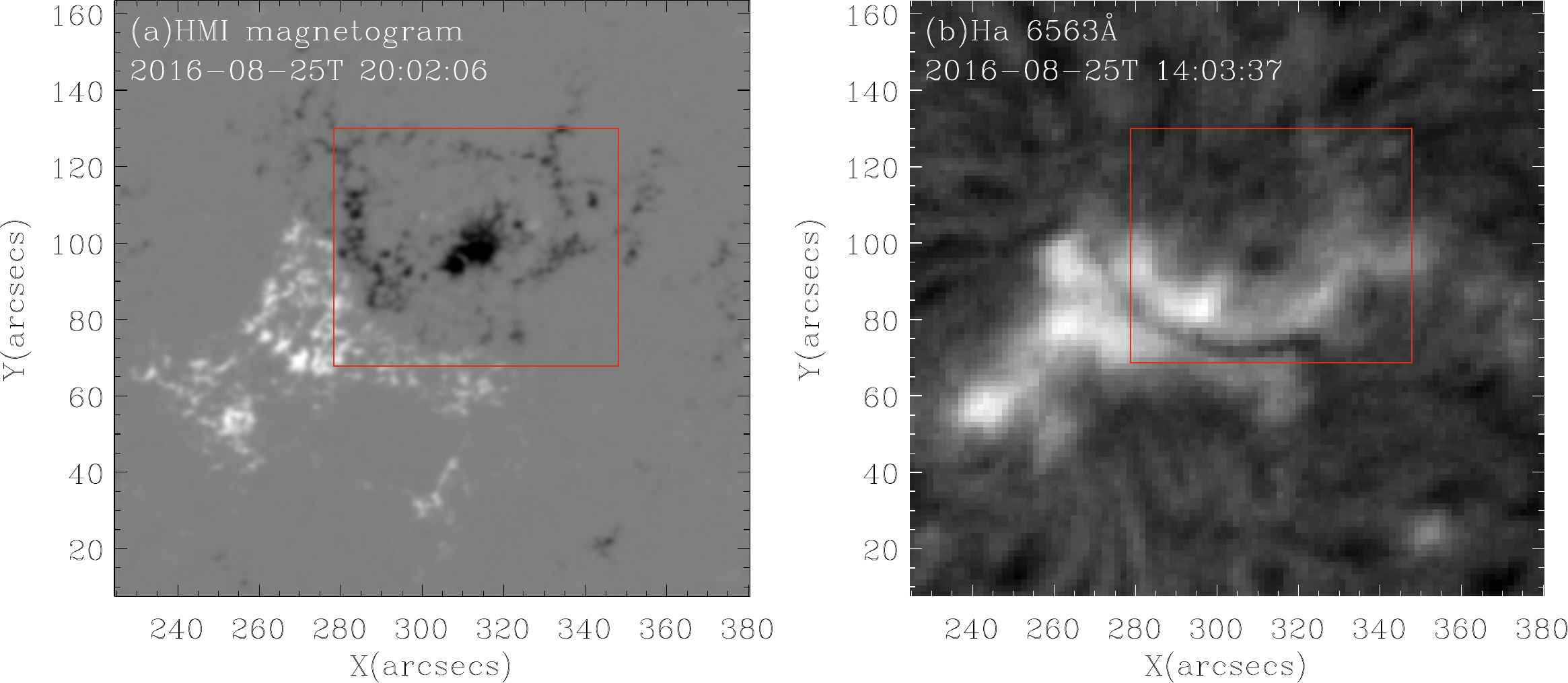}
\includegraphics[width=180mm,angle=0,clip]{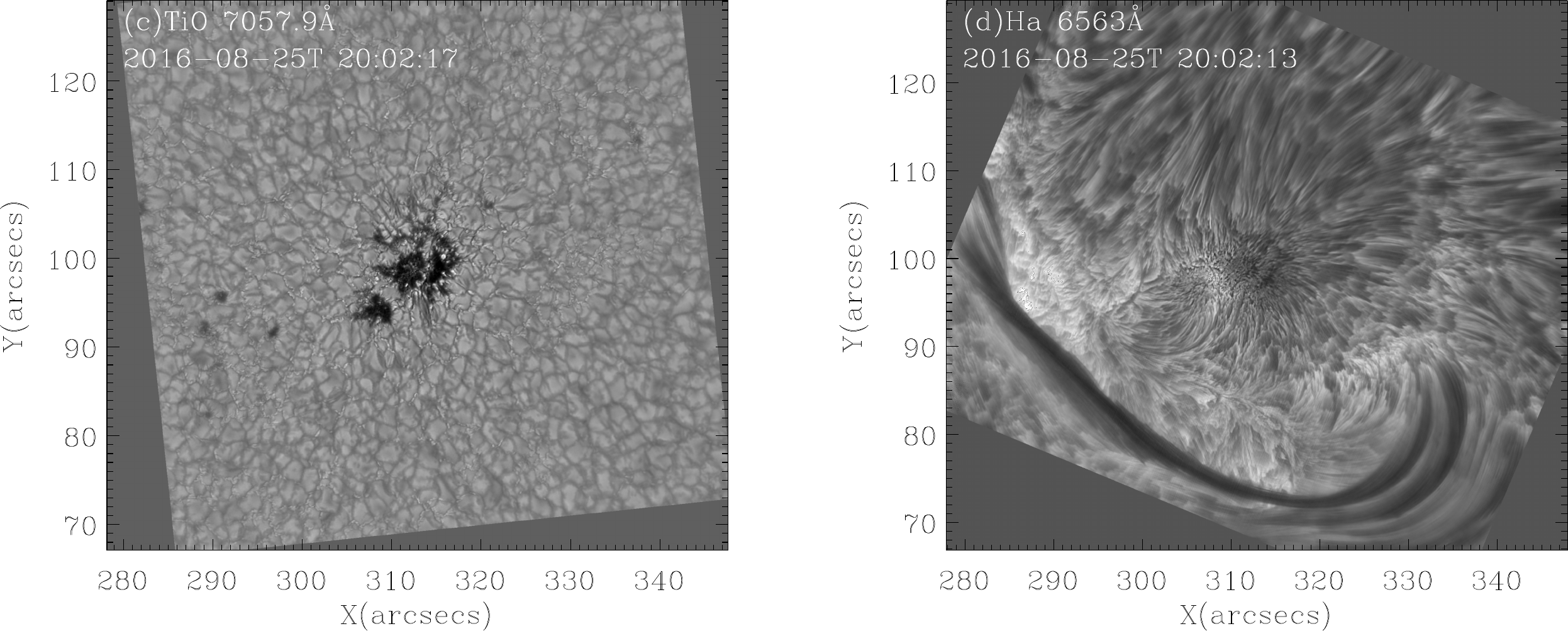}
\end{minipage}

\caption{Active region NOAA 12579 on August 25  2016: HMI magnetogram (panel a). H$\alpha$ image (Bass2000 survey -panel b). GST images of the leading sunspot in TiO (panel c), in H$\alpha$ (panel d). The red boxes in panel a and panel b have the same FoV, corresponding to the FoV of panel c and panel d.
}
\label{figure_1}
\end{figure*}

\begin{figure*}[!htbp]
\begin{minipage}{\textwidth}
\centering
\includegraphics[width=180mm,angle=0,clip]{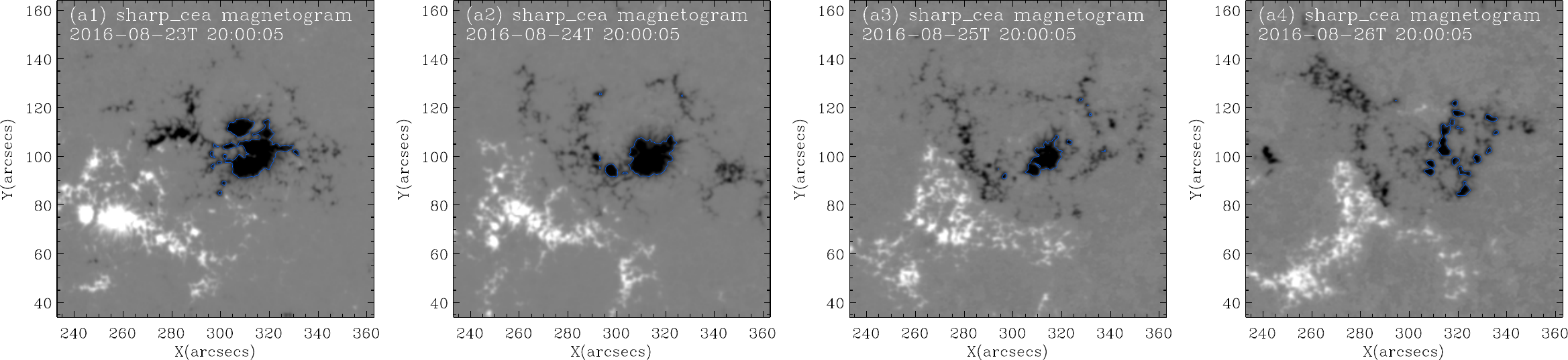}
\includegraphics[width=180mm,angle=0,clip]{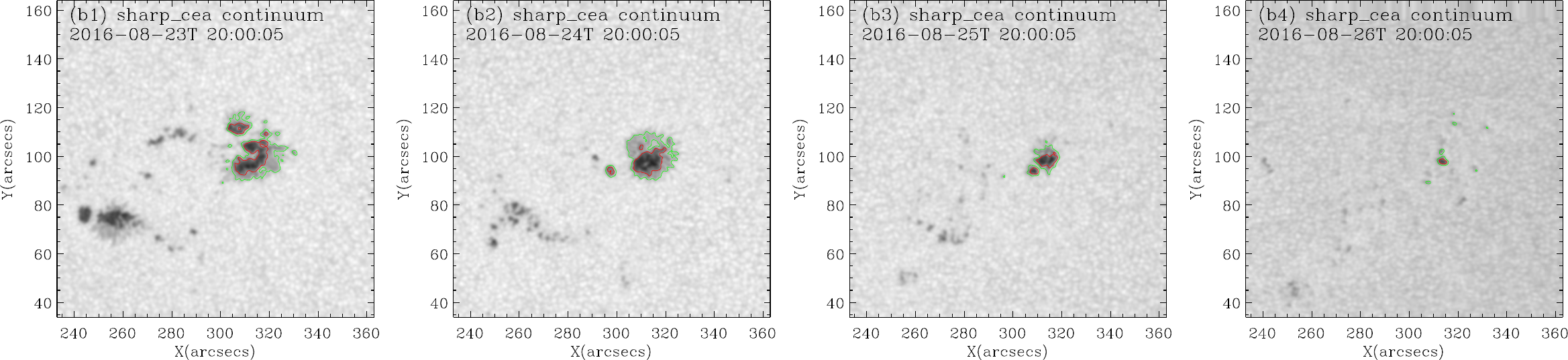}
\end{minipage}

\caption{Temporal evolution of 
the active region NOAA 12579 observed with the SDO/HMI between August 23 and August 26, 2016: {\it Top panels:} Magnetograms, with the blue contour lines representing the magnetic field boundary of -600 G. {\it Bottom panels:} Continuum images, with the red, and green contour lines corresponding to umbra and penumbra values of 0.7 $I_0$ and 0.87 $I_0$, respectively.}
\label{figure_2}
\end{figure*}

\begin{figure*}[!htbp]
\begin{minipage}{\textwidth}
\centering
\includegraphics[width=180mm,angle=0,clip]{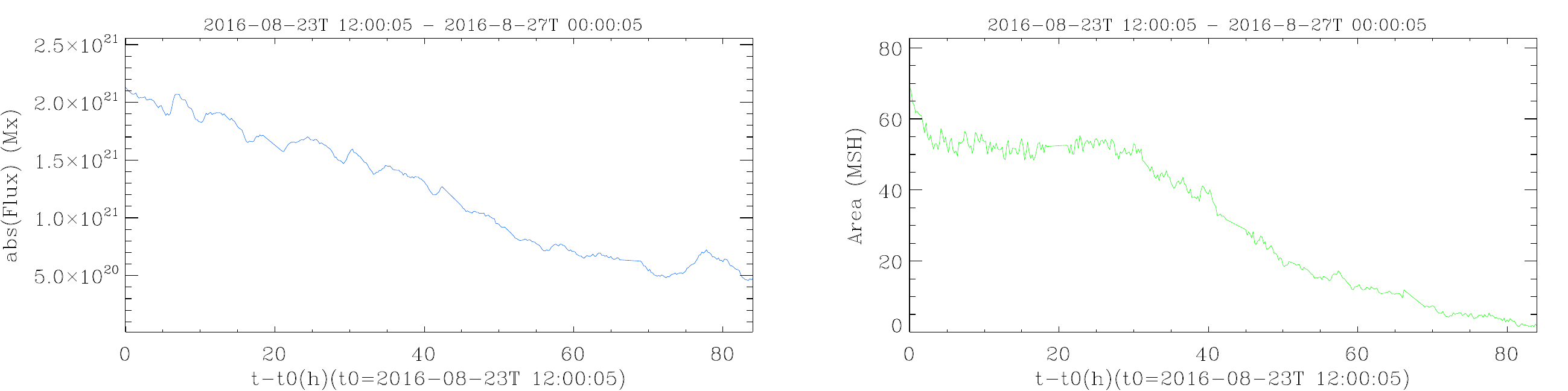}
\end{minipage}

\caption{Variation in the magnetic flux corresponding to the blue contour in Fig.~\ref{figure_2} during the decay taking place between August 23, 2016, 12:00 UT and August 27, 2016, 00:00 UT (left).  Variation in the area corresponding to the green contour in Fig.~\ref{figure_2} during  decay between August 23, 2016 12:00 UT and August 27, 2016 00:00 UT (right).}
\label{figure_3}
\end{figure*}

\begin{figure*}[!htbp]
\begin{minipage}{\textwidth}
\centering
\includegraphics[width=150mm,angle=0,clip]{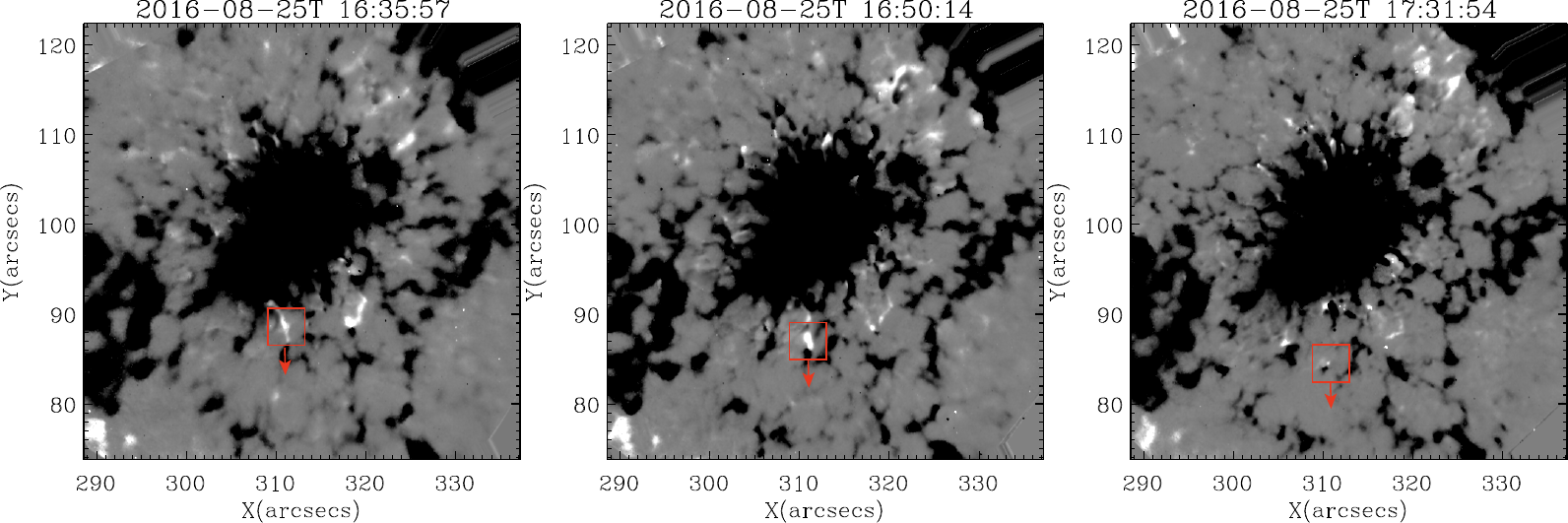}
\includegraphics[width=150mm,angle=0,clip]{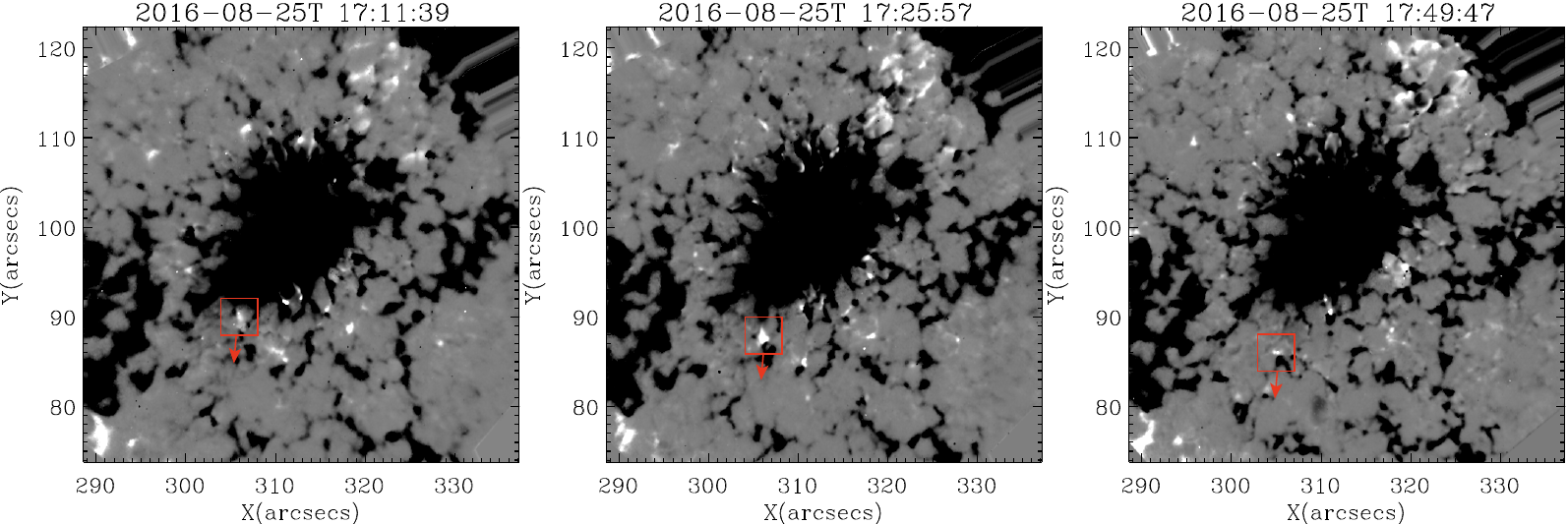}
\includegraphics[width=150mm,angle=0,clip]{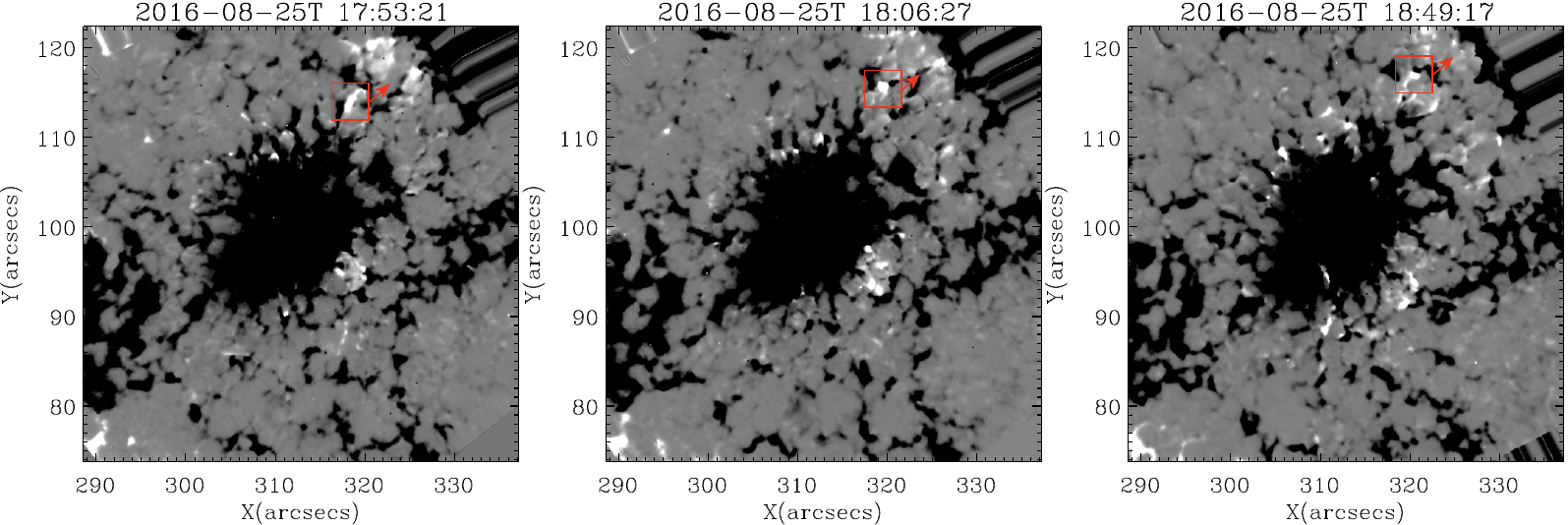}
\end{minipage}
\caption{
Longitudinal magnetic field evolution observed with the GST/NIRIS spectrograph for three MMFs marked in red boxes escaping away from the sunspot in the moat region. The arrows indicate the direction of their trajectory. In each raw a MMF is followed for three times. The black  lines in the diagonal are artifacts due to the field rotation. }
\label{figure_4}
\end{figure*}

\section{Introduction}

Sunspots are the most straightforward phenomenon to observe in studies of solar activity, as the core and obvious signature of active regions. Sunspots exhibit a complex structure with a  dark area in the center, called the umbra, and a light-dark area surrounding the umbra, called the penumbra. Studying the decay process of sunspots is very important for understanding the physical mechanism of sunspots.

A significant part of the magnetic field covering the solar surface comes from the destruction of sunspots, especially near the maximum of activity. The decay of active regions (ARs) results in the dispersion of their magnetic field which dominates the large-scale structure visible on the Sun \citep{VanDriel1998}.  Commonly it is accepted that 
the photosphere flows are responsible for the advection of magnetic flux, redistribution of flux during the decay of sunspots, and build-up of magnetic shear in flaring active regions \citep{Verma2011}. 
The flux emergence rate on AR scales is eight times higher at solar maximum than at minimum \citep{VanDriel1998}. The active region evolution is asymmetric with an emergence lasting around five days and a decaying phase between 70\% to 94\% of their entire lifetime  (several months) \citep{harvey1993}. Two modes of sunspot decay exist, fast mode fragmentation and gradual decay. 

In this work, we are most interested in the second mode, where the decay of a sunspot is a slow process. Many studies have focused on sunspot erosion and the dispersion of the magnetic field generated by sunspots \citep{verma2012}. An annular outflow called the moat flow \citep{Sheeley1969} is observed at the surface, close to the sunspot. The moat flow is a large-scale flow pattern commonly observed around sunspots \citep{meyer1974}. However, flux removal and dispersal can only be understood in the context of the moat flow's fine structure. Moving magnetic features (MMFs) are bright features that correspond to small magnetic elements of mixed polarity travelling radially outwards while immersed in the granulation surrounding sunspots \citep{Sheeley1972};  \citep{Harvey1973}; \citep{Hagenaar2005}. In brief, MMFs move radially outward with a velocity of 1 km s$^{-1}$, during sunspot decay,  before they reach and dissolve within the network, at the boundaries of the supergranular cell containing the sunspot \citep{verma2012}. Thus, in the vicinity of a decaying sunspot, most of the G-band bright points (GBPs), which are believed to be associated with thin magnetic flux tubes, appear to be born close to a magnetized plasma. Previous studies have been concentrated on motions of the 
GBPs relative to the families of granules on short sequence duration (2h) \citep{roudier2003}. Families of repeatedly splitting granules are present in the sunspot moat. These families of granules are advected by the outward flow in the moat \citep{bonet2005}. The radial motions confine the GBPs within the channels with the same average speeds as the neighboring granules. These “passively” moving GBPs are carried along by the same large-scale flows as families \citep{bonet2005}. Here, we take the benefit of a longer time sequence (6h) with magnetic field measurement to study the evolution of the magnetic elements through the moat and families of granules.

The decay of sunspots area is an important symbol of sunspot decay. \citet{bumba1963} studied that the phase of area decrease essentially progresses at two rates: rapidly and slowly. The rapid one is in small spots in the central and following regions of the group, and the slow one is in regular spots. In addition to the decay of area, the decrease in magnetic flux is also an important process. By calculating the variation of magnetic flux with time, it is found that the magnetic flux of sunspots decreases linearly with time \citep{skumanich1994,verma2012,sheeley2017}. 

\citet{Rempel_sub2011} developed numerical simulations of active regions based on flux emergence and confirmed the decrease in the flux during the decay phase \citep{rempel2015}.
 The first process invoked for the decay of sunspot is the diffusion by turbulence.
The spot formation phase transitions directly into a decay phase. Subsurface flows fragment the magnetic field and lead to intrusions of almost field-free plasma underneath the photosphere. When such intrusions reach photospheric layers, the spot fragments. The timescale for spot decay is comparable to the longest convective timescales present in the simulation domain. They found that the dispersal of flux from a simulated spot in the first two days of the decay phase is consistent with self-similar decay by turbulent diffusion.
The dispersion of sunspot flux is also visible in the surrounding of the spot with the moat region with up and down flows.
The absence of downflows perturbs the upflow and downflow mass flux balance and leads to a large-scale radially overturning flow system \citep{rempel2015}.

Another proposed mechanism is based on the short life of  MMFs.
They are mostly seen during the decay phase of sunspots \citep{li2019}. \citet{shine2001} distinguished among three types of MMF. Type I are the classical bipolar features that are  the most representative of MMF activity. Type II are unipolar features with the same polarity as the spot. Finally, type III refers to unipolar features of opposite polarity to the sunspot. Types I and III would be related to the extension of the penumbra filaments beyond the sunspot's outer boundary somehow. \citet{Pillet2002} proposed that MMFs as a continuation of sunspots penumbra. They also argued that MMFs originated from the interaction of the field-free convection (moat flows) and the Evershed magnetized channels outside the spot.

\citet{roudier2016} proposed that the trees of fragmenting granules (TFGs), also referred to as families of granules, appear as one of the major elements of the supergranules that diffuse and advect the magnetic field on the Sun’s surface. The strongest supergranules contribute the most to magnetic flux diffusion in the solar photosphere . The largest (but not the most numerous) families are related to the strongest flows and could play a major role in supergranule and magnetic network formation \citep{Malherbe2018}. Exploding granules constitute the strongest horizontal flows on the quiet Sun and contribute to the structure of the surface horizontal velocity fields that build the large-scale organization of the discrete magnetic field \citep{roudier2020}.


In this paper, we present high-resolution observations of a decaying sunspot obtained by the  1.6 meter  Goode Solar Telescope (GST) operating at the Big Bear Solar Observatory (BBSO), as well as by the Helioseismic and Magnetic Imager (HMI) aboard the Solar Dynamic Observatory (SDO). We consider the dynamic evolution of the decaying sunspot and the magnetograms in Section~2. We analyze the relationship between families of granules,
mesogranules, and photospheric networks in Section~3. We introduce the methods we used in Section~4. We analyze the families motion linked to radial horizontal flows and corks diffusion in Section~5. We analyze the families proper motions and magnetic elements diffusion in Section~6. In Section~7, we summarize our results regarding the cause of the decaying sunspots.


\begin{figure*}[!htbp]
\centering
\includegraphics[width=16cm]{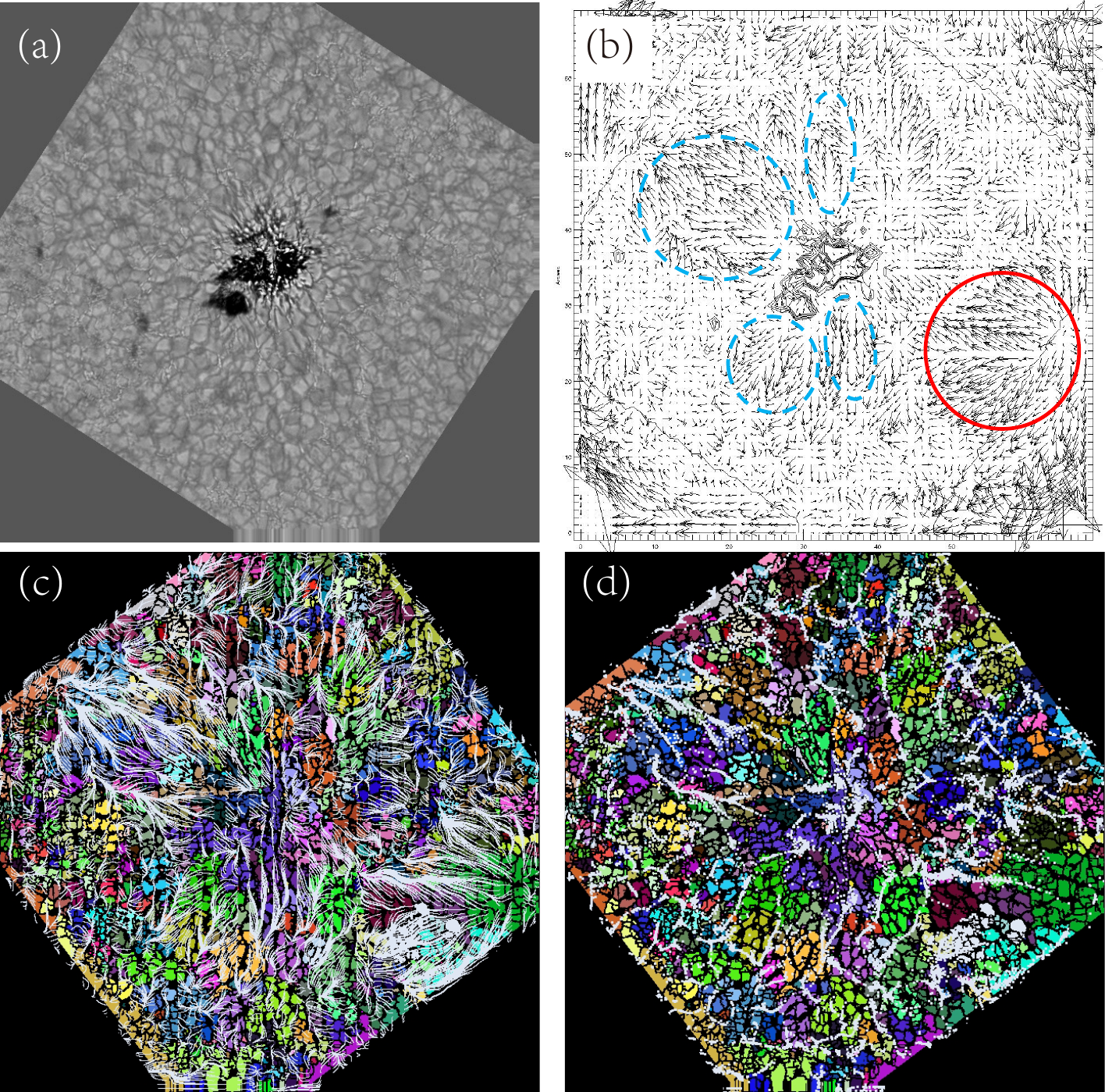}
\includegraphics[width=8cm]{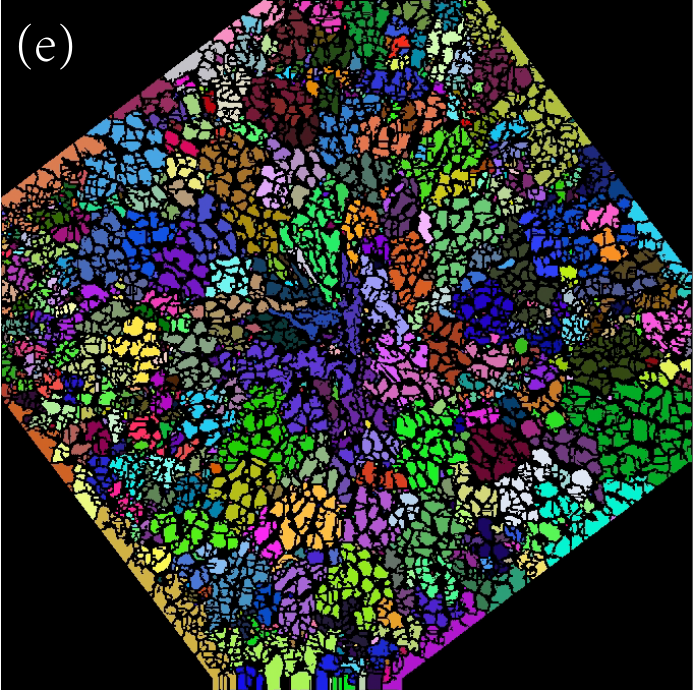}
\caption{ Zoom on the leading spot of AR 12579 observed with GST in TiO in panel a.  
Panel b shows the horizontal velocities averaged over all the time sequence. The amplitude is between 0  and 2 km s$^{-1}$. The contour of the  sunspot umbra (black line) is visible in the center of the image.  An example of supergranule flow is surrounded by the red line. Blue dashed line contours show high radial moat flows going away from the sunspot. Panel c:\ TFGs (or families of granules)   obtained with the horizontal velocies after  5001 secs (at 17:58:33 UT) overlapped by cork trajectories, each  TFG has a  specific  color (blue, red, green...). Panel d: Same  TFG map  at the same time 
overlapped by  the cork location. Panel e: Map of the TFGs  alone  at 17:58:33 UT in  the different colors.
We note that the cork locations are  dispersed toward  the edges of the TFGs. 
The field of view is  $70\arcsec\times 70\arcsec$. }
\label{velo}
\end{figure*}


\begin{figure}[!htbp]
\includegraphics[width=8cm]{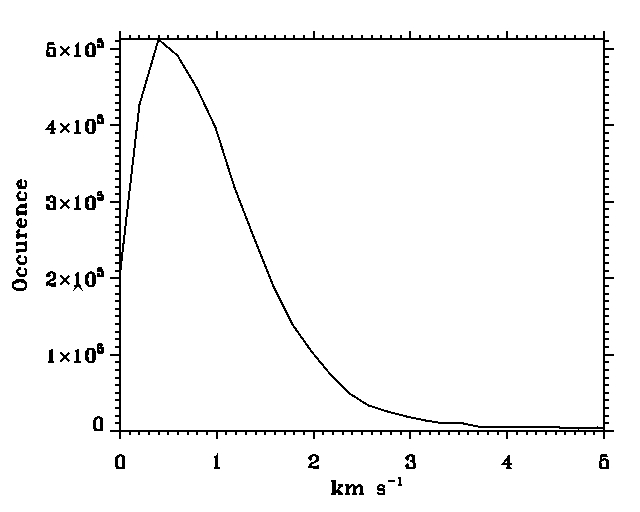}
\caption{Horizontal velocity histogram  obtained from the local correlation tracking method (LCT).}
\label{histo}
\end{figure}

\begin{figure*}[!htbp]
\centering
\includegraphics[width=16cm]{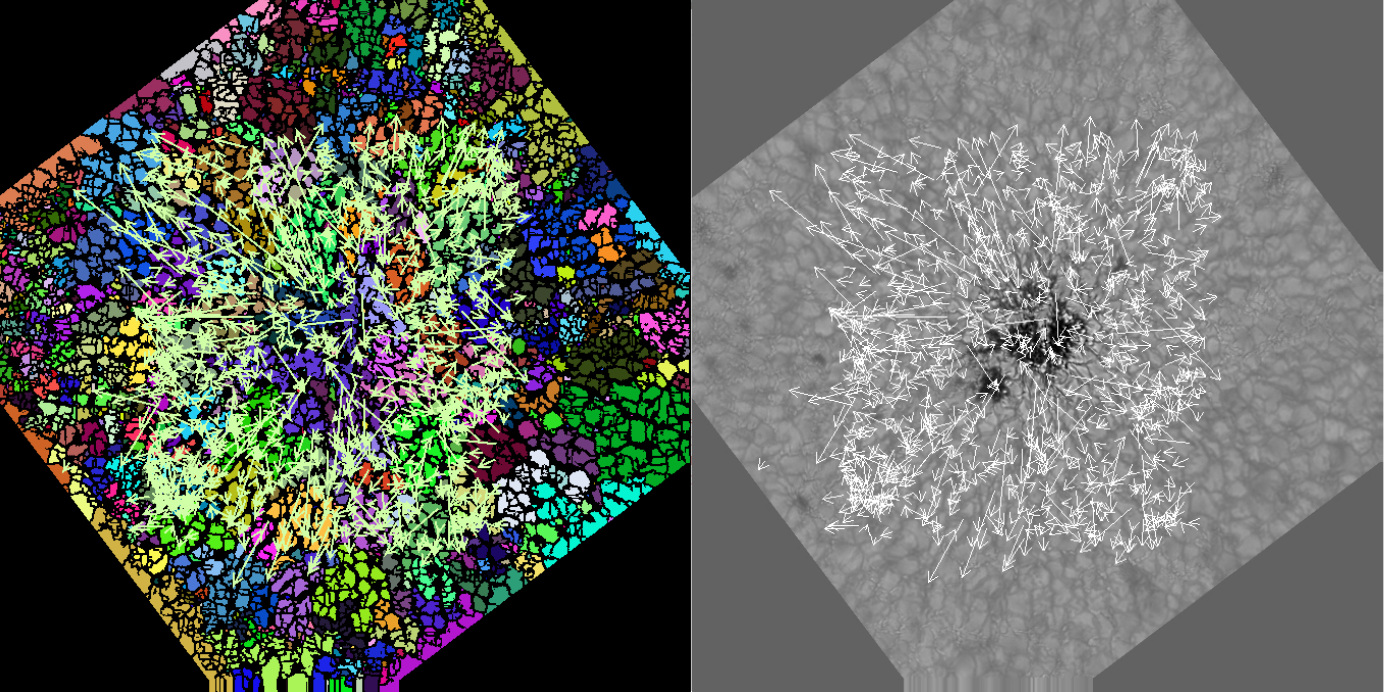}
\caption{Large TFG radial trajectories (hell green arrows) over the granule families (left panel), the same trajectories (white arrows) over the sunspot observed in TiO (right panel).  The FoV is $70\arcsec\times 70\arcsec$.}
\label{TFG3}
\end{figure*}


\begin{figure}[!htbp]
\includegraphics[width=9cm]{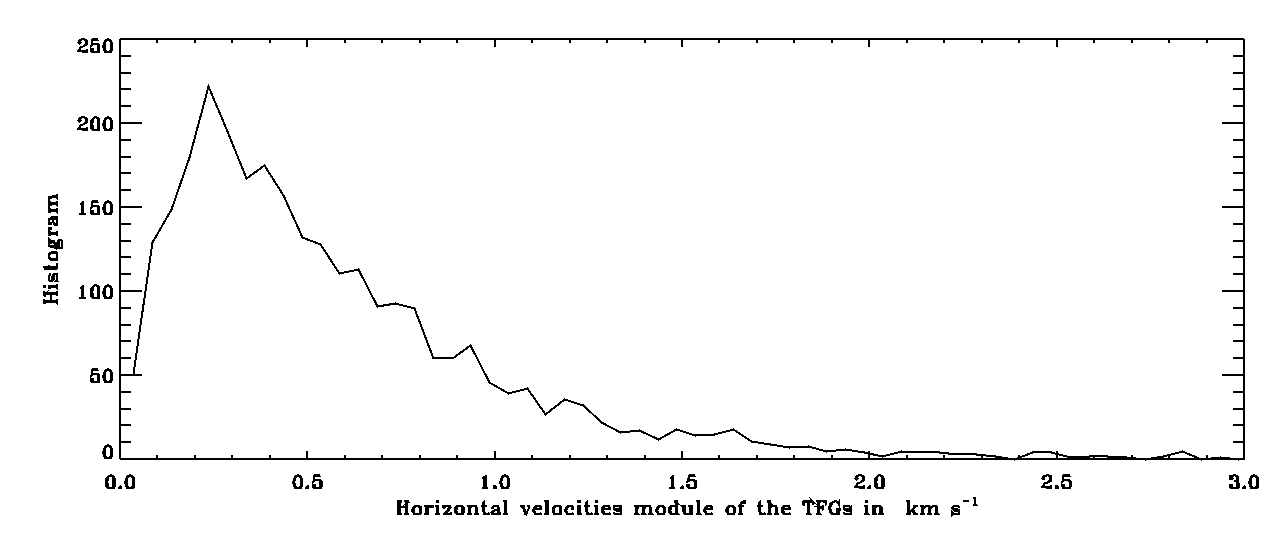}
\caption{Large TFG velocities (radial) histogram}
\label{figure 8}
\end{figure}

\begin{figure*}[!htbp]
\includegraphics[width=18cm]{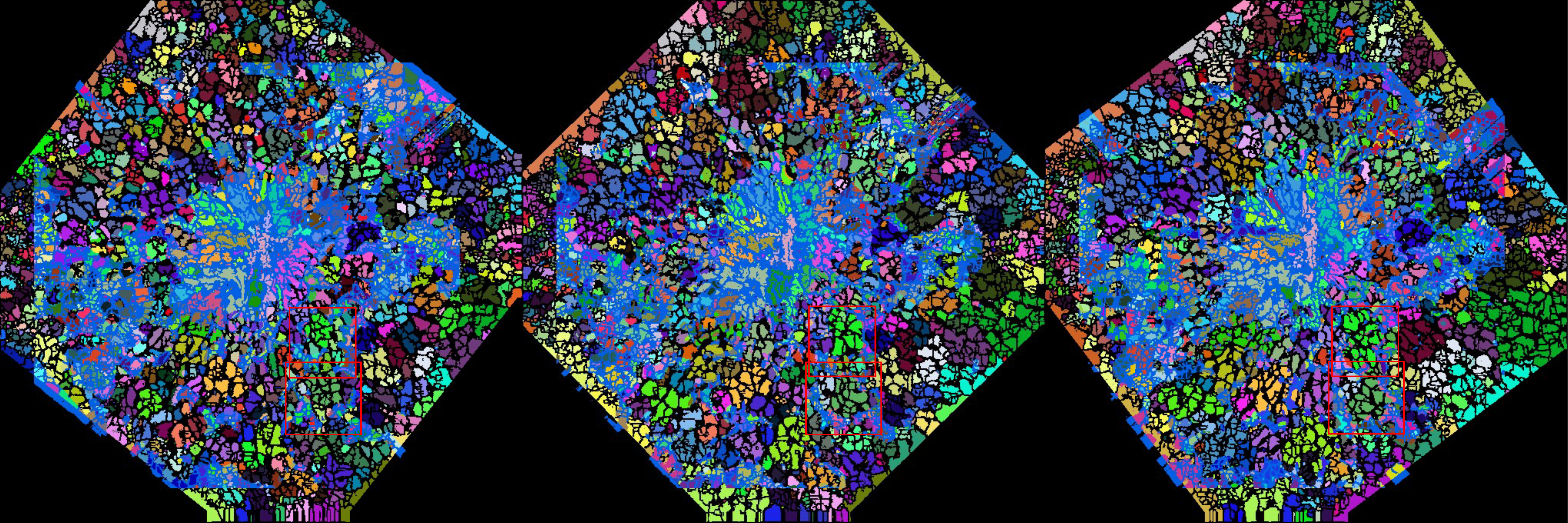}\\
\includegraphics[width=18cm]{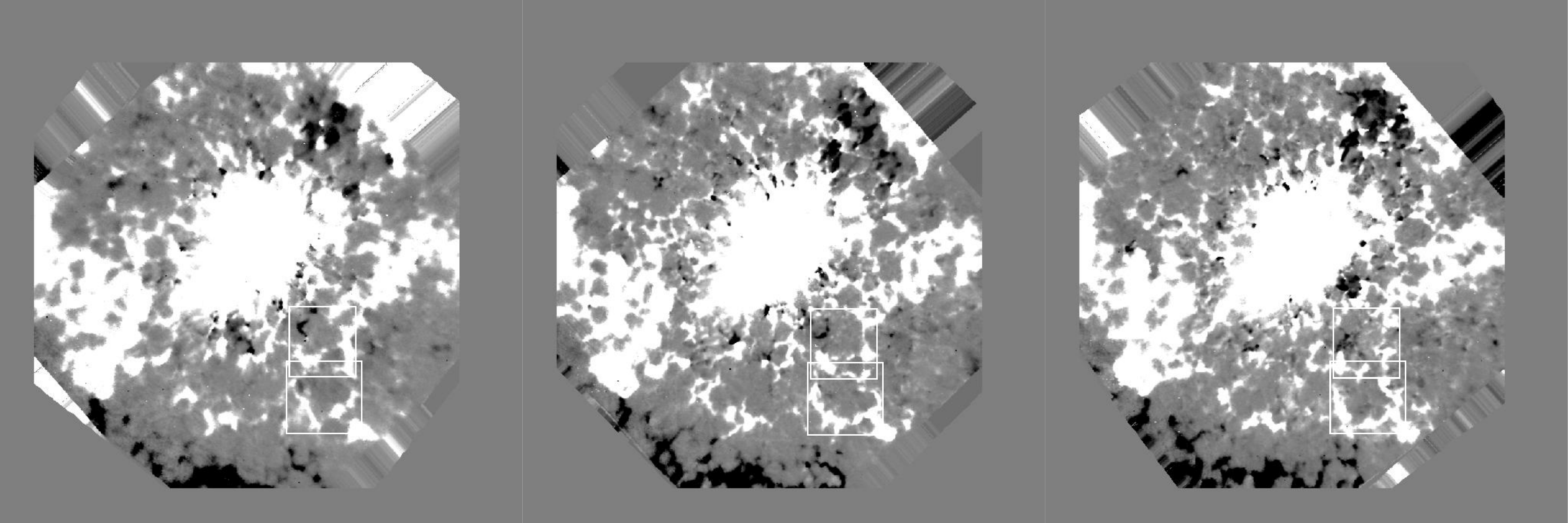}
\caption{TFG evolution at three  different times  (top)  following the granules with the horizontal flows after t= 1427  secs (at 16:58:59 UT), after t= 3214 secs (at 17:28:48 UT), and after t= 5001 secs (at 17:58:33 UT) and superimposed NIRIS magnetic field (blue). Bottom: Evolution of the magnetic field (now in white). The magnetic field has been reversed for a better view of the MMFs. The two squares locate the TFG1 and TFG2 examples shown in more detail in  Figs.~\ref{TFG1} and~\ref{TFG2}. The FoV is  $70\arcsec\times 70\arcsec$ and the threshold of the magnetic field is 20 Gauss. The black or white  lines in the diagonal are artifacts due to field rotation. Animation of the magnetic field  in blue is provided in movie Intb-paper.mp4. }
\label{TFGA}
\end{figure*}

\begin{figure}[!htbp]
\includegraphics[width=8cm]{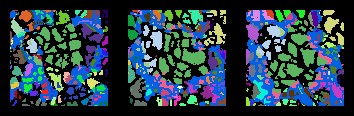}\\
\includegraphics[width=8cm]{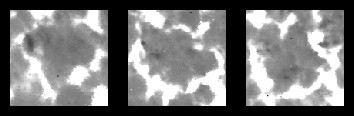}
\caption{ Zoom on TFG 1 (green color) evolution at three different times (top)  following the granules with the horizontal flows after t= 1427  secs (at 16:58:59 UT), after t= 3214 secs (at 17:28:48 UT), and after t= 5001 secs (at 17:58:33 UT) 
superimposed NIRIS  magnetic field (blue). Bottom: Evolution of the magnetic field network  (now in white) showing the global motion of the network. For example the   round shape  network  cell is approaching the bottom of the image due to   the  "displacement"  of TFG1 towards the bottom. The magnetic field has been reversed for a better view of the MMFs. The FoV is  $9.9\arcsec\times 9.7\arcsec$.}
\label{TFG1}
\end{figure}

\begin{figure}[!htbp]
\includegraphics[width=7.9cm]{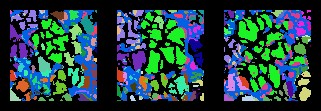}\\
\includegraphics[width=7.9cm]{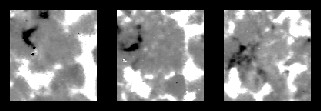}
\caption{  Zoom of TFG 2 (green color) evolution at different times (top)  following the granules with the horizontal flows after t= 1427  secs (at 16:58:59 UT), after t= 3214 secs (at 17:28:48 UT), and after t= 5001 secs (at 17:58:33 UT) 
superimposed  NIRIS magnetic field (blue). Bottom: Evolution of the magnetic field network (now in white)  showing the network approaching the bottom part of the panel due to  "the expansion and magnification"  of TFG2 towards the bottom.  The magnetic field has been reversed for a better view of the MMFs. Field of view  $8.8\arcsec\times 9.2\arcsec$. } 
\label{TFG2}
\end{figure}

\begin{figure*}[!htbp]
\begin{minipage}{\textwidth}
\centering
\includegraphics[width=180mm,angle=0,clip]{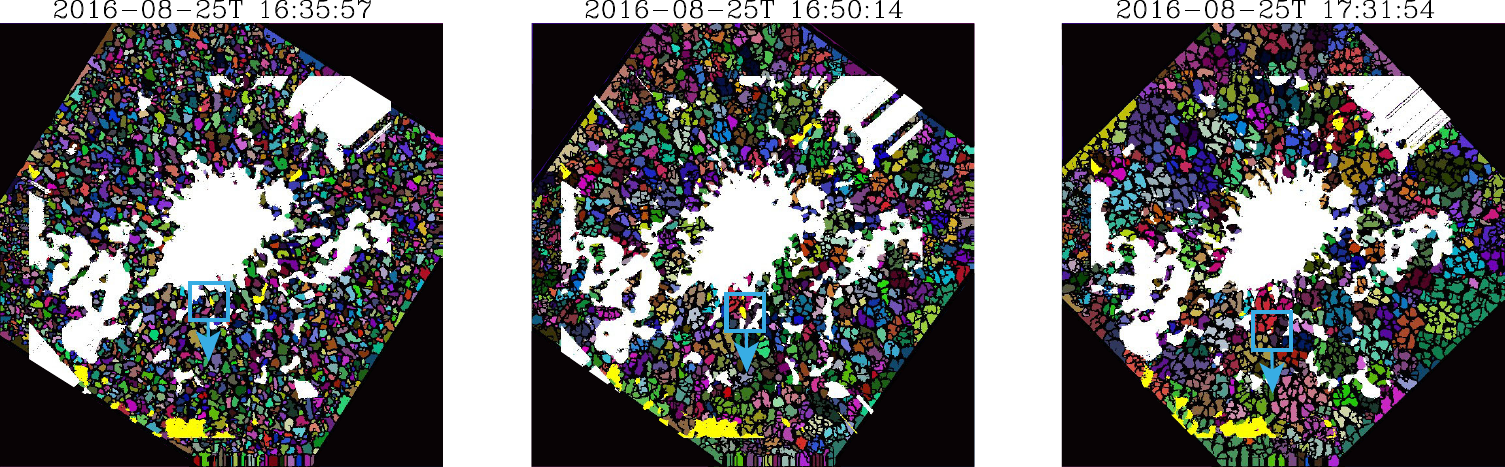}
\includegraphics[width=180mm,angle=0,clip]{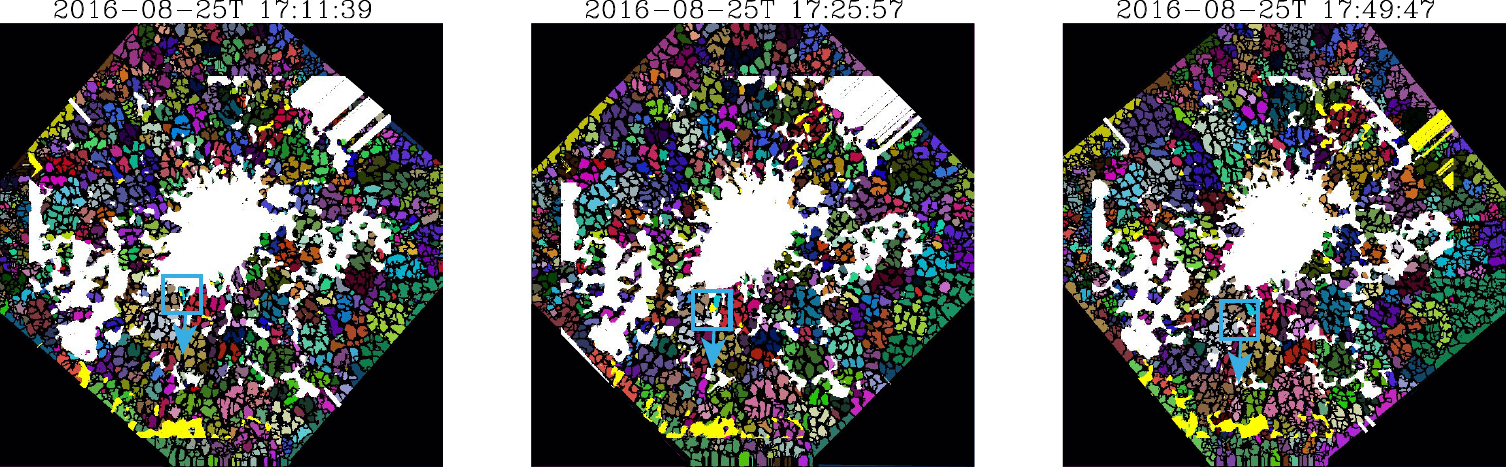}
\includegraphics[width=180mm,angle=0,clip]{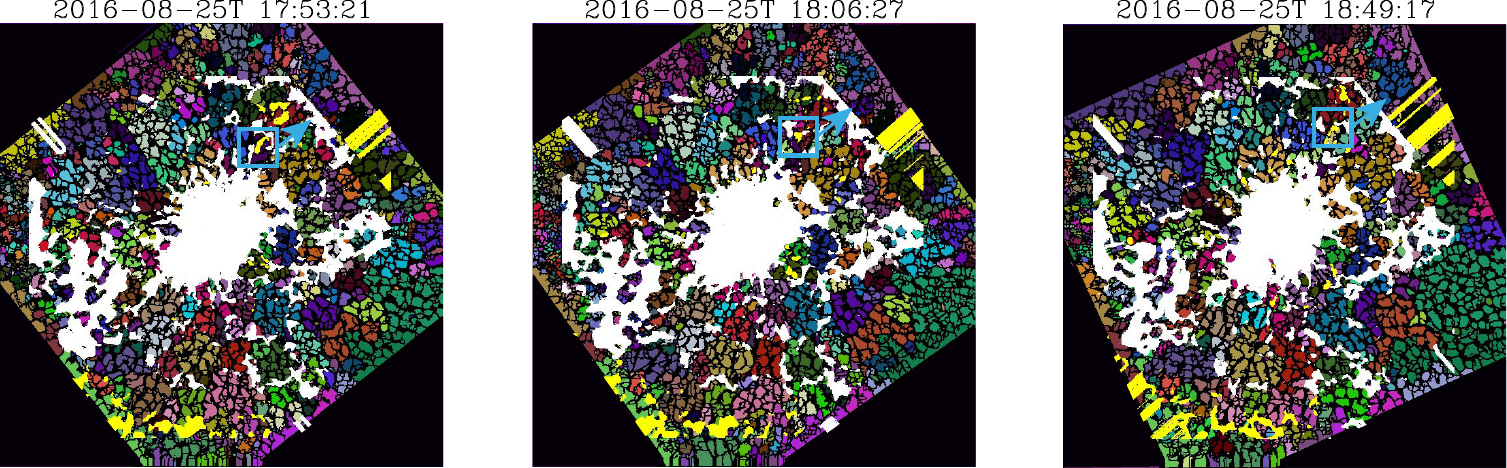}
\end{minipage}
\caption{Example of the evolution of  the three  MMFs marked in cyan boxes  visible in Figure~\ref{figure_4} (red boxes) for three different times in TFG  images  superimposed NIRIS magnetic field (in yellow and white).  Each row follows  a  differenet MMF.  In each cyan box, there a white small area that indicates the presence of magnetic field or MMF.
The arrows indicate the direction of their trajectories as
in Figure~\ref{figure_4}. The threshold of the magnetic field is 20 Gauss. The panels of this  figure are extracted from the movie (Famb.mp4).   The white  or  yellow  lines in the diagonal are artifacts due to field rotation.}
\label{figure_5}
\end{figure*}

\section{ Observations}
\subsection{Instruments}

On August 25, 2016, between 16:35:12 UT  and 22:32:44 UT (duration 5h57min32sec)  the  GST 
was observing the leading sunspot in the NOAA active region (AR) 12579 located at N11W22.  
We focused on the sunspot in a field of view around 70$^{\prime\prime}$×70$^{\prime\prime}$ using the TiO line. We also used  the Solar Dynamic Observatory (SDO) \citep{pesnell2012} coupled with the Helioseismic and Magnetic Imager (HMI) \citep{scherrer2012,Schou2012}  

The GST data contain simultaneous observations of the photosphere, using the titanium oxide (TiO) line taken with the Broadband Filter Imager (BFI) \citep{Cao2010}, the passband of the TiO filter is 10 \AA{}, centered at 705.7~nm, while its temporal resolution is about 15 s with a pixel scale of 0.$^{\prime\prime}$034. We obtained a full Stokes spectroscopic polarimetry using the Fe I 1565~nm doublet over a 58$^{\prime\prime}$ round field of view (FoV) with the aid of a dual  Fabry-P{\'e}rot etalon by the NIRIS Spectropolarimeter \citep{cao2012}. The Stokes I, Q, U, and V profiles were obtained every 72 s with a pixel scale of 0.$^{\prime\prime}$081. All TiO data were speckle reconstructed using the Kiepenheuer-Institute Speckle Interferometry Package \citep{Woger2008}.

We analyze the continuum intensity data by the Helioseismic and Magnetic Imager (HMI) \citep{Schou2012} on board the SDO spacecraft. Generally, HMI provides four main types of data: dopplergrams (maps of solar surface velocity), continuum filtergrams (broad-wavelength photographs of the solar photosphere), and both the line-of-sight and vector magnetograms (maps of the photospheric magnetic field). The processed hmi.sharp\_cea\_720s continuum intensity and magnetogram data were obtained with a 12~\rm{min} cadence and a 0.03$^{\circ}$  pixel size, which were corrected by projection and provided by the HMI team. Continuum intensity maps of HMI help us to co-align the TiO images and the magnetograms taken by GST. The GST images taken at each wavelength position were internally aligned using the cross-correlation technique provided by the BBSO programmers. The co-alignment between the SDO/HMI continuum and GST images was achieved by comparing commonly observed features of sunspots in FeI 6173 \AA{} images and TiO images taken frame by frame.
The GST images taken at each wavelength position were internally aligned using the cross-correlation technique provided by the BBSO programmers.

\subsection{Decaying sunspots}
\label{The decaying sunspot}

Figure \ref{figure_1} presents a HMI magnetogram of the full active region NOAA 12579 and an image in H$\alpha$ obtained by the Meudon survery (BASS2000.com).  A  zoom of the leading spot in magnetogram and in H$\alpha$ which is provided by  GST observations. We note that 
on August 25 2016, the leading sunspot umbra is still concentrated in  the HMI magnetogram while  in the  TiO images, we observe a strong fragmentation  of the spot with light bridge. In H$\alpha$  the fibrils are nearly all radial and a large filament is observed along the inversion polarity line. It will be interesting to analyse the anchorage of the fibrils around the sunspot versus the fragmentation of the sunspot.

In the present work, we are interested in the process of the decay of the leading negative sunspot of AR 12579, which occurred over  several days.
Therefore, it was necessary  to analyse a long series of observations from August 23 to August 26  2016. HMI  provides this long series of magnetograms and continuum images  (Figure \ref{figure_2}).
To provide a 
quantitative representation of the decaying sunspot, we calculated the total magnetic flux and the area decay respectively in  the sunspot limited by the blue and green contours 
over a period of 3.5 days between August 23 2016 12:00 UT and August 27 2016 00:00 UT (Figure~\ref{figure_3}). {From the graph, it is evident that the magnetic flux exhibits an  exponential 
decrease from 2.13 × $10^{21}$ Mx to approximately 4.58 × $10^{20}$ Mx.  }



In order to accurately delineate the umbra and penumbra of the sunspot, we computed the average quiet sun continuum intensity near the sunspot, ($I_0$), the umbra was defined as the region with intensities below 0.7$I_0$ ($I_u$<0.7$I_0$), while the penumbra was characterized as the area with intensities greater than 0.7$I_0$ but less than 0.87$I_0$ (0.7$I_0$<$I_p$<0.87$I_0$). In Figure~\ref{figure_2}, the umbra is outlined in red, and the penumbra is outlined in green. At the commencement of the observation, a portion of the sunspot's penumbra had dissipated, and by the conclusion of the observation, the penumbra of the sunspot had nearly entirely vanished. From the continuum intensity images, a light bridge has emerged in the center of the sunspot. The appearance of  light bridges indicates the onset of the umbra’s fragmentation \citep{li2021}.

Figure~\ref{figure_3} displays the evolution of the area decay of the studied sunspot region, indicating an exponential  reduction in area. The plateau of the curve can be explained by the process of fragmentation which occurs not continuously but by blocks. The area unit is expressed in millionths of a solar hemisphere (MSH), where 1 MSH is equivalent to 3.321 Mm$^{2}$.
The decaying sunspot's area decreased from 99.256 MSH to 2.507 MSH. These findings confirm that the sunspot is in a decaying phase.  The curves of Figure \ref{figure_3} are in agreement with the laws found by \citet{Petrovay1997,Petrovay1999}.

\subsection{MMFs in the moat of pores}
\label{MMFs in the moat of pores}

The following work has been obtained by the high temporal and spatial resolution of the GST telescope. 
We provide a  movie for understanding the decay phase of the sunspot and the formation of the TFGs.
The movie\footnote{movie\_paper.mp4} shows  three panels: the left shows  the evolution of the continuum intensity in TiO observed with the BFI, the middle shows the  longitudinal magnetic field obtained with NIRIS,  and the right shows the formation of the  TFGs throughout the whole the sequence (see Section 3). Snapshots of  the central panel are shown in Figure \ref{figure_4}, the left panel in Figure \ref{velo}a, the  right panel in Figure \ref{velo} e. We note that the FoV of the GST observations is rotating and we made the choice  to keep the sunspot in the center  always oriented towards the north. Artifacts were observed in the diagonals, which we could not remove\footnote{We  also retained all the data and this is not convenient for looking at the movie with random bad images. We apologise for this inconvenience.}. \\

 High-resolution observations with the NST reveal ultrafine magnetic loop structures originating from compact magnetic flux concentrations in the intergranular lanes on the surface and reaching up to the solar corona \citep{Ji2012}. Small flux tubes are transported away by the large-scale supergranulation from a sunspot, which are seen as MMFs in the moat \citep{li2019}. Figure~\ref{figure_4} shows the magnetograms of the decaying sunspot at nine times acquired by GST/NIRIS. It can be observed that certain moving magnetic features (MMFs) are in the moats of pores. They exhibit monopolar characteristics with polarity opposite to the sunspot, thereby classifying them as Type III MMFs. In this area \citet{jin2022}, found a new form of flux appearance, namely, magnetic outbreak, in the hecto-Gauss region of pore moats. The rapid emergence, explosion, and final dissipation constitute the whole process of magnetic outbreak. We selected 22 MMFs and outlined three of them in red boxes in Figure~\ref{figure_4}.
  From Table~\ref{Table}, we can see that the mean lifetime of the MMFs is 54.4 min. The horizontal velocity is between 0.27 km s$^{-1}$-1.49 km s$^{-1}$. \citet{2005ApJ...635..659H} found that the magnetic features have a lifetime of 1 hr, which is broadly consistent with the conclusions we have drawn here. From Figure~\ref{figure_4} and Table~\ref{Table}, it can be observed that during the extended four-hour observation period, the MMFs radially moved outward from the dark penumbral fibrils continually.

\citet{Pillet2002} consolidated opinion in the sense that MMF activity is the observed signature of sunspot decay through diffusion. A rough calculation (note:\ a more accurate would give a higher rate) indicates that the average creation rate for MMFs is 5.5 MMFs h$^{-1}$. According to \citet{shine2001}, the channels were produced by several MMFs moving along these paths. This number of channels requires an MMF creation rate of 5 MMFs h$^{-1}$ or higher. In summary, we find that MMFs can contribute to the sunspot evolution.


\begin{table}[h]
\setlength{\tabcolsep}{3pt}
\caption{Characteristics for  a number of long-duration MMFs detected in the moat region}
\label{Table} 
\centering 
\begin{center}
\begin{tabular}{c c c c c} 
\hline\hline 
Event & Appearing time  & Ending time  &Lifetime & Horizontal velocity  \\ 
 &(UT)&(UT)&(min)& (km s$^{-1}$)\\
\hline 
1 & before 16:35 & 17:31 &  & 0.58 \\ 
2 & 16:59 & 17:58 & 59 & 0.27 \\
3 & 17:10 & 17:53 & 43 & 0.91 \\
4 & 17:50 & 19:05 & 75 & 0.54 \\
5 & 18:01 & 18:56 & 55 & 0.64 \\
6 & 18:08 & 18:53 & 45 & 1.49 \\
7 & 18:12 & 18:43 & 31 & 0.29 \\
8 & 18:17 & 19:30 & 73 & 0.51 \\
9 & 18:30 & 19:05 & 35 & 0.71\\
10 & 18:31 & 19:17 & 46 & 0.25 \\
11 & 18:37 & 19:22 & 45 & 0.71 \\
12 & 19:30 & 20:27 & 57 & 0.65 \\
13 & 19:50 & 21:16 & 86 & 0.76 \\
14 & 20:09 & 21:27 & 78 & 0.67 \\
15 & 20:18 & 20:59 & 41 & 0.66 \\
16 & 20:20 & 21:34 & 74 & 0.42 \\
17 & 20:45 & 20:38 & 53 & 0.94 \\
18 & 21:18 & 21:47 & 29 & 1.23 \\
19 & 21:25 & after22:31 &  & 1.17 \\
20 & 21:26 & after22:31 &  & 0.74 \\
21 & 21:27 & after22:31 &  & 0.70 \\
22 & 22:00 & after22:31 &  & 0.98 \\

\hline 
\end{tabular}
\end{center}
\end{table}

\section{TFGs and MMFs}
\subsection{Relationship between families of granules, mesogranules, 
and the photospheric network}

One of the main goals in solar physics is to understand the formation and decay of sunspots on the solar surface:\ how magnetic flux formed in the convective zone is then distributed, advected, and diffused over the solar surface \citep{Sheeley2005,Malherbe2015}.
The analysis of a solar granulation sequence obtained at the Pic du Midi Observatory showed that the convective cells of such granules exhibit an organized evolution by forming families of granules \citep{Roudier2004}. Those authors 
confirm the existence of TFGs. Thus, TFGs
seem to play a role in the diffusion of the magnetic elements on the Sun's surface
\citep{Roudier2004}. Therefore, these results demonstrate that the long-living families contribute to the formation of the magnetic network and suggest that supergranulation could be an emergent length scale building up as small magnetic elements are advected and concentrated by TFG flows \citep{Roudier2009}.

The evolution of TFGs and their mutual interactions result in cumulative effects that are able to build horizontal coherent flows with a longer lifetime than granulation (1 to 2 hours) over a scale of up to 12 arc secs. These flows clearly act on the diffusion of the intranetwork (IN) magnetic elements and also on the location and shape of the network. From the analysis (lasting 24 hours), TFGs appear as one of the major elements of the supergranules which diffuse and advect the magnetic field on the Sun’s surface. The strongest supergranules contribute the most to magnetic flux diffusion in the solar photosphere \citep{roudier2016}. They found that TFGs and horizontal surface flows (provided by the LCT) can be detected either from intensities or Vz/Vdop components, for high-resolution observations and numerical simulations. They apply this method to a 3D run providing the Vz component in depth. This reveals a close relationship between surface TFGs and vertical downflows 15 Mm below the surface \citep{roudier2019}. Finally, families of repeatedly splitting granules are present in the sunspot moat. The motions in the moat of long-lived TFGs are radially orientated drift of centroids away from the spot \citep{bonet2005} and could contribute to the decay of sunspots.


\subsection{Method of tracking the granules}

A TFG consists of a family of repeatedly splitting granules, originating from a single granule at its beginning \citep{roudier2003}. TFGs are used in this study as a tool to quantify the temporal and spatial organization of solar granulation at large scales.
The long-lived families then tend to control the long-time evolution of the corks distribution. Although the time series of the sunspot  in the  previous study of \citet{bonet2005} was too short to form well-defined TFG.
Here, with the long series of GST TiO observations,  we 
detected the beginning phase of their formation; for that purpose, granules were labeled in time according to the method described in \citet{roudier2003}. The TFGs were detected after oscillation filtering of intensities using a segmentation and labelling technique. Surface horizontal flows were derived from local correlation tracking (LCT) \citep{November1988} and from  the intensities or magnetic flux elements. 
From our previous analysis (lasting 24 hours), TFGs appear as one of the major elements of the supergranules that
diffuse and advect the magnetic field on the Sun’s surface. The strongest supergranules contribute the most to magnetic
flux diffusion in the solar photosphere \citep{roudier2016}.

\subsection{Families motion linked to radial horizontal Flows and corks diffusion}

The TiO observation sequence (5h57min) of the active region AR 12579 located near the disk center,  with the measurement of the magnetic field, offers the possibility of studying the TFGs motions relative to the sunspot center and their interaction with the flux tubes escaping from the sunspot. As the first step, in order to gain an overview of the movements around the sunspot, we used the LCT method to determine the horizontal velocities throughout this sequence. The results are shown in Figure \ref{velo} (panel b),
where we can see radial motions around the sunspot at the center of the image. A component of a large supergranule is also visible in the bottom right of the figure (red circle). Most of the velocities are up to 2 km s$^{-1}$, with a peak of 0.4 km s$^{-1}$(Fig.~\ref{histo}), which is generally measured at the solar surface.



 One way to know more about the role of TFGs in the diffusion of magnetic elements is to compute the advection of floating corks by the granular flow. This technique was already used by \citep{Roudier2004}. It consists of following the trajectories of floating corks, initially uniformly distributed, and characterising their spatial distribution after some time.
Figure \ref{velo} (panel c) shows the cork's trajectories during the first 5001 seconds of the sequence, which are radially directed away from the sunspot on the edge of the TFGs in majority. Figure \ref{velo} (panel d) 
reveals the final cork locations after a diffusion by the horizontal flows described above. From this figure, we observe that corks are located mainly around the TFGs, 
which confirms the important role of the TFGs in the diffusion of the magnetic field. These corks are proxies of the magnetic field flux tube, which seem a good approximation of the evolution of the magnetic flux relative to TGFs evolution and motions.
The evolution of the TFGs alone  (snapshot in Fig. \ref{velo} e) is shown in movie paper (right panel).  

 \subsection{Families proper motions  and  magnetic elements diffusion}

Thanks to our time sequence of the sunspot with the measurement of the longitudinal magnetic field, we have been able, for the first time, to analyze the behavior and interactions between the TFGs and the flux tubes escaping from the sunspot during its decay.  The evolution of the displacement of the center of gravity of each TFG has been measured. Fig.~\ref{TFG3} shows the families of granules and superimposed the arrows (red) indicating the exact displacements of the TFGs with an area larger than 100 Mm$^{2}$ throughout their motions, to allow a clear plot. In the great majority, the TFGs move radially from the sunspot center over several Megameters. The histogram of the velocities of their center of gravity peaks around 0.22 km s$^{-1}$, with some up to 3 km s$^{-1}$  (Fig.~\ref{figure 8}).

 This observation of the radial motions of the TFGs from the sunspot center enables us to go on and study the behavior of the magnetic field flux tubes of the network. The movie\footnote{Int-paper.mp4} shows the evolution of the magnetic field in blue overlaid to TiO images.
 
 At the top of  Fig.~\ref{TFGA}, we follow the evolution of the TFGs at the times between t=1427 and 5001 seconds, which correspond to  16:58:59 UT and 17:58:33 UT in our sequence, on which the magnetic field has been superimposed (in blue). The magnetic field is clearly located at the edge of the TFGs.  In particular, the distribution and diffusion of the magnetic field are closely linked to the evolution of the TFGs. For example, 
 a detailed inspection of the two areas surrounded by squares in  Fig.~\ref{TFGA}  clearly shows the evolution of the TFGs and the magnetic field as they move radially away from the center of the sunspot (Fig.~\ref{TFG1}  and Fig.~\ref{TFG2}).  In Fig.~\ref{TFG1}, the magnetic white network is approaching  the low boundary of the panel due to the displacement of the green TFG,  while  in  Fig.~\ref{TFG2} it is due to the expansion and magnification of the green TFG.  The magnetic field tubes of the network  are subject to the motions and expansion of the TFGs.

 More precisely, we find two modes of action for the diffusion of the magnetic field towards the exterior of the sunspot as a function of the evolution of the TFGs. The first mode is visible in Fig.~\ref{TFG1} which shows an example of the displacement of a TFG (green: radial velocity of 0.354 km s$^{-1}$) with the magnetic network (blue) around it. We can clearly see that the TFG's own movement pushes the magnetic field outwards from the sunspot.  
The second mode is  related to the radial expansion of the TFGs; for instance, in Fig.~\ref{TFG2},  it is visible where the TFGs only grows in one direction (velocity 0.202 km s$^{-1}$).  In the present case, it travels radially, which drags the magnetic field along its edges with it.
  We measure in both cases the radial velocity of the magnetic flux tube on the TFGs border lying between 0.1 to 1.8 km s$^{-1}$. The last high-velocity amplitude is due to the expansion of the granules at the edges of the magnetic flux tubes. 

\subsection{Relationship between MMFs and TFGs}

 The evolution of the TFGs overlaid with the magnetic field is shown in the movie\footnote{Famb.mp4} and  as a snapshot in  Figure \ref{figure_5}.
Figure \ref{figure_5}  presents the family of granules for the same times as the Figure \ref{figure_4}. The red boxes indicate the
MMFs. We note that in the middle of the box, there are white points indicating the presence of magnetic flux between color pixels corresponding to TFGs. There is a good correspondence between MMFs and flux tubes escaping from the sunspot.

\section{Conclusion}

In this paper, we report on coordinated observations obtained  with the GST at BBSO  (BFI imager and NIRIS spectro polarimeter) 
of the active region NOAA 12579 on August 25, 2016. The observed event shows the process of a  decaying sunspot according to SDO/HMI.
 We took advantage of the high temporal and spatial resolution of the GST  instruments (TiO images and NIRIS magnetograms) that has allowed us to follow magnetic flux tubes in the moat region
interacting with the families of granules.  The  long sequence of observations (5h57min) allows  us to quantify the temporal and spatial organization of the granulation at large scales in families (TFGs). 

We obtained the following results:
\begin{enumerate}
 \item We calculated the total magnetic flux and area within the entire sunspot. These all exhibited an approximately linear decline, while the magnetic flux decreased from 2.13 × $10^{21}$ Mx to 4.58 × $10^{20}$ Mx. The area reduced from 99.256 MSH to 2.597 MSH. These findings confirm that the sunspot is in a decaying phase.
 
 \item  We found 22 MMFs in the moats of pores. They continually moved outward from the dark penumbra fibrils for four hours and their mean lifetime is 54.4min. MMF activity is a signature of sunspot decay through diffusion and can contribute to the sunspot evolution.
 
 \item  We followed the emergence of granules inside the sunspot border. The magnetic fields that are expelled from the sunspot are located at the edge of the TFGs.

\item We chose TFGs that were well surrounded by a magnetic field to calculate horizontal velocities. The magnetic field at the border of each TFG is taken away from the sunspot at a larger horizontal velocity due to the development (magnification or displacement) of the area of the TFGs. The velocities of the flux tubes could be accelerated due to the expansion of  TFGs reaching  3 km s$^{-1}$. The mean value of their displacements (Table 1) is of the order of the moat flow   that is commonly measured with lower spatial resolution observations.
\end{enumerate}
In conclusion, this detailed study indicates that the TFGs and their development contribute to diffuse the magnetic field outside the sunspot.
 The relationship between GBPs and MMFs has not been treated in this study because GBP is considered as a magnetic field proxy. In this paper,  we  directly focus on studying the magnetic field tubes. 
It will  be subsequently useful to test their relationship  based on  our high-resolution observations (beyond the scope of the present paper). 
 The relationship of the TFGs with the chromosphere pattern has not been studied either.
In Figure \ref{figure_1} we show an H$\alpha$ image showing the penumbra and the fibrils that are not visible in the TiO image.  It would be interesting to understand the coupling between the flow along the fibrils (Evershed flow) and the convection flow, but this is again out of the scope of this paper. In a future paper, this aspect will be studied, along with the poorly understood proces of penumbra fibril formation. 
All these questions remain open and ripe for future study.




\


\begin{acknowledgements}
Guiping Ruan and Chenxi Zheng acknowledge the support by the NNSFC grant 12173022, 11790303 and 11973031. We gratefully acknowledge the use of data from the Goode Solar Telescope (GST) of the Big Bear Solar Observatory (BBSO). BBSO operation is supported by US NSF AGS-2309939 and AGS-1821294 grants and New Jersey Institute of Technology. GST operation is partly supported by the Korea Astronomy and Space Science Institute and the Seoul National University. W. Cao acknowledges support from US NSF AST-2108235 and AGS-2309939 grants. We thank Chunlan Jin and Jingxiu Wang formulated the data observation of BBSO/GST.
B.S.\ acknowledge support from the European Union’s Horizon 2020 research and innovation programme under grant agreement No 870405 (EUHFORIA 2.0) and the ESA project "Heliospheric modelling techniques“ (Contract No. 4000133080/20/NL/CRS). This work was granted access to the HPC resources of CALMIP under the allocation 2011-[P1115]. These works are supported by COFFIES, NASA Grant 80NSSC20K0602. We thank SDO/HMI, SDO/AIA teams for the free access to the data.  
 
\end{acknowledgements}


\bibliography{reference.bib}
\bibliographystyle{aa}

\end{document}